\documentclass[lettersize,journal]{IEEEtran}
\usepackage{amsmath,amsfonts}
\usepackage{algorithmic}
\usepackage{algorithm}
\usepackage{array}
\usepackage[caption=false,font=normalsize,labelfont=sf,textfont=sf]{subfig}
\usepackage{textcomp}
\usepackage{stfloats}
\usepackage{url}
\usepackage{verbatim}
\usepackage{graphicx}
\usepackage{cite}

\usepackage{BOONDOX-cal}
\usepackage{amssymb}
\usepackage{amsmath}

\usepackage{array}
\usepackage{tabularx}
\usepackage{multirow}
\usepackage{float}
\usepackage{tabularx}

\usepackage[T1]{fontenc}
\usepackage[sort&compress]{natbib}
\usepackage{booktabs}

\usepackage{xcolor}

\newtheorem{lemma}{\bf Lemma}
\newtheorem{theorem}{\bf Theorem}
\newtheorem{assumption}{\bf Assumption}
\newtheorem{remark}{\bf Remark}
\newtheorem{proposition}{\bf Proposition}

\begin{document}
	
	\title{Stochastic Tube-based Model Predictive Control for Cyber-Physical Systems under False Data Injection Attacks with Bounded Probability}
	
	\author{Yuzhou Xiao, Senchun Chai,~\IEEEmembership{Senior Member,~IEEE,} Li Dai, Yuanqing Xia,~\IEEEmembership{Fellow,~IEEE,}\\ and Runqi Chai,~\IEEEmembership{Senior Member,~IEEE}

	\textit{This article has been accepted for publication in the IEEE Transactions on Systems, Man, and Cybernetics: Systems. Copyright may be transferred without notice, after which this version may no longer be accessible.}

	\thanks{Yuzhou Xiao, Senchun Chai, Li Dai, Yuanqing Xia, and
Runqi Chai are with the School of Automation, Beijing Institute of
Technology, Beijing 100081, China 
}
	\thanks{Corresponding author: Runqi Chai}
}

	\maketitle
\begingroup\renewcommand\thefootnote{}\footnote{© 2025 IEEE. Personal use of this material is permitted. Permission from IEEE must be obtained for all other uses, in any current or future media, including reprinting/republishing this material for advertising or promotional purposes, creating new collective works, for resale or redistribution to servers or lists, or reuse of any copyrighted component of this work in other works.}\endgroup

	\begin{abstract}
	This paper addresses the challenge of amplitude-unbounded false data injection (FDI) attacks targeting the sensor-to-controller (S-C) channel in cyber-physical systems (CPSs). We introduce a resilient tube-based model predictive control (MPC) scheme. This scheme incorporates a threshold-based attack detector and a control sequence buffer to enhance system security. We mathematically model the common FDI attacks and derive the maximum duration of such attacks based on the hypothesis testing principle. Following this, the minimum feasible sequence length of the control sequence buffer is obtained. The system is proven to remain input-to-state stable (ISS) under bounded external disturbances and amplitude-unbounded FDI attacks. Moreover, the feasible region under this scenario is provided in this paper. Finally, the proposed algorithm is validated by numerical simulations and shows superior control performance compared to the existing methods.
	\end{abstract}
	
	
	\section{Introduction}

The rise of the Internet of Things has necessitated the integration of traditional industrial control systems into networks, giving birth to cyber-physical systems (CPSs) \cite{IoT}\cite{Cao2024}. CPSs enable cloud-based monitoring and control of edge subsystems \cite{CPS1}\cite{CPS2}\cite{HSun2021}. However, this connectivity also exposes systems to vulnerabilities and network attacks \cite{security}\cite{Wu2021}, such as denial-of-service (DoS) and false data injection (FDI) attacks, the latter being a common form of deception \cite{attacks}\cite{Pang2022}.
FDI attacks pose significant threats to CPSs. In power grids, these attacks manipulate sensor data, leading to false readings and incorrect control decisions \cite{De2024}. Such attacks can cause power outages, equipment damage, and large-scale blackouts, undermining grid stability and reliability \cite{Wu2024}. Similarly, in unmanned aerial vehicles, FDI attacks alter navigation data, causing deviations from intended paths or even crashes \cite{Li2024}. These risks threaten mission success and endanger people and property. The disruptive potential of FDI attacks underscores the urgent need for resilient security measures to detect and mitigate such threats.
To counter these risks, an offense-defense dynamic emerges between attackers and defenders in protecting CPSs \cite{game}\cite{Cai2023}.

The sensor-controller (S-C) channel is a critical vulnerability point in CPSs, as it transmits sensor data to the controller. The resilient model predictive control (MPC), which is one of the popular resilient schemes, relies heavily on current system measurements, making the S-C channel's integrity crucial for maintaining optimality and accurate system recognition. While much of the literature has focused on the controller-actuator (C-A) channel \cite{Franze_severe}\cite{Yang2023}\cite{He2022}, the S-C channel remains underexplored. Addressing this gap is critical, as it directly affects the reliability of the system's decision-making process. Our methods can also complement existing defenses to provide a more comprehensive security approach.

This paper focuses on a CPS with actuator input saturation and physical constraints on system states, presenting the challenge of managing both state and control input constraints while ensuring resilience against FDI attacks. MPC is a key method for addressing such multi-constraint problems, offering unique advantages \cite{MPCconstraints}. MPC's rolling optimization characteristics help mitigate the impact of temporary open-loop situations, making it effective against DoS and severe deception attacks \cite{Franze_severe}\cite{Sunqi_DoS}.

However, traditional robust MPC schemes, including min-max MPC, show poor performance under cyber attacks \cite{Sunqi_phd}. To address this, tube-based MPC leverages disturbance-invariant sets, ensuring the disturbed system remains close to the nominal system \cite{Mayne2005}. While effective against bounded disturbances and amplitude-bounded FDI attacks, this algorithm faces limitations with higher-level attacks. Therefore, it is essential to develop resilient MPC methods that can detect, estimate, and compensate for the effects of these attacks, particularly amplitude-unbounded FDI attacks, while considering the resource limitations and constraints of the CPS.

As defenders in control systems, our primary responsibility is to protect the system's last line of defense. This entails implementing resilient strategies to mitigate the impact of attacks when the system is inevitably compromised by network threats \cite{resilient}\cite{Yao2021}. 
This paper presents a resilient methodology for detecting a particularly harmful class of cyberattacks, namely the so-called amplitude-unbounded FDI attacks. The proposed approach is specifically designed for CPSs vulnerable to such modeled FDI attacks on the S-C channel. The remainder of this section provides an overview of existing studies, followed by a detailed description of our specific contributions.

\subsection{Related studies and key differences}

Recent research in resilient control strategies for CPSs has focused on countering vulnerabilities to cyber attacks, particularly FDI attacks. Several approaches, including resilient MPC, have been proposed to enhance system resilience during attacks (see, for instance, \cite{Franze_severe}\cite{Yang2023}\cite{Sunqi_DoS}\cite{Eventtrigger}\cite{He2023}).

Based on traditional tube-based MPC methods \cite{Mayne2005}\cite{MayneE2000}, which focus on additive disturbances, these recent studies combine predictive and reactive strategies to more effectively mitigate the impact of cyber attacks on CPSs.
For example, \cite{Sunqi_DoS} utilizes control buffers to maintain stability during temporary open-loop situations. \cite{Yang2023} provides guarantees of robust constraint satisfaction and uniformly ultimately bounded behavior, offering a less conservative and more effective solution for attack mitigation. \cite{He2023} addresses network resource limitations by introducing a self-triggered strategy, using a signal reconstruction mechanism to prioritize critical control data and facilitate control sequence recovery after an attack. \cite{MFPFC2024} presents a model-free predictive control method that removes the need for state estimation or system modeling, offering greater flexibility in managing communication disruptions.

Nevertheless, the effectiveness of resilient MPC critically hinges on its attack detection capabilities, presenting several unresolved challenges. First, existing frameworks often rely on restrictive assumptions regarding attack duration and channel refreshing \cite{Franze_severe}\cite{Sunqi_phd}, which fail to capture the dynamic and evolving nature of modern cyber threats. This detection-dependent conservatism potentially compromises real-time resilience when detection is delayed or inaccurate. Second, balancing detection accuracy with computational/communication overhead remains problematic for large-scale CPSs. Many methods incur significant resource costs during scalability \cite{Franze_severe}\cite{Eventtrigger}, increasing vulnerability to resource-draining false positives \cite{He2023}. Unlike observer-based techniques \cite{RobustFilter}, our approach embeds state buffering directly within the MPC formulation to reduce detection-induced computational redundancy. Finally, current strategies exhibit limited adaptability to heterogeneous attack patterns, as the efficacy of mitigation mechanisms (e.g., \cite{Yang2023}\cite{He2023}) is inherently constrained by the detection scope and its capability to distinguish attack types.

The various features achieved in different studies are summarized in Table \ref{Comparisons}.
\begin{table}[h!]
	\centering
	\renewcommand\arraystretch{1.2} 
	\caption{Comparisons of the Published Papers and This Paper}
	\setlength{\tabcolsep}{1.3mm}{
		\begin{tabular}{c|c|c|c|c|c|c|c}
			\toprule[2pt]
			\makebox[0.1cm][c]{Function} & \makebox[0.1cm][c]{\cite{Franze_severe}} & \makebox[0.1cm][c]{\cite{Yang2023}}
			& \makebox[0.1cm][c]{\cite{Sunqi_phd}} & \makebox[0.1cm][c]{\cite{Eventtrigger}} &  \makebox[0.1cm][c]{\cite{He2023}} & \makebox[0.1cm][c]{\cite{MFPFC2024}} &\makebox[1cm][c]{This paper}    \\
			\toprule[1pt]
			Unbounded attack resilience & Yes & Yes & No & No & No & No & Yes \\
			Strong robustness & No & Yes & No & No & No & Yes & Yes \\
			Nonlinear scalability & Yes & No & Yes & Yes & Yes & No & Yes \\
		Robustness to false positives & Yes & No & No & No & No & No & Yes \\
			Optimality guarantee & Yes & Yes & No & Yes & No & No & Yes \\
			\toprule[2pt]
		\end{tabular}
		\label{Comparisons}}
\end{table}

\subsection{Contributions}

The challenge of integrating detection and mitigation for amplitude-unbounded FDI attacks in CPSs remains largely underexplored. While some works use mechanisms like zero-order hold \cite{ReviewDetection}\cite{LiDropout}, our approach reconstructs the control signal using a feasible solution from tube-based MPC. This method's effectiveness hinges on the sufficiency of a pre-stored control sequence and the persistent feasibility of the optimization problem, which are central to our contributions. A key feature of our work is the integration of a threshold-based attack detector with a tube-based MPC control sequence buffer. This combination ensures system stability against both bounded disturbances and amplitude-unbounded FDI attacks. We also introduce a novel activation mechanism to minimize false positives and use a probability model to validate channel refreshing, avoiding restrictive assumptions. Numerical simulations confirm our method enhances CPS resilience, expands the feasible operating region, and outperforms related techniques, demonstrating its practical effectiveness.

In summary, this paper makes the following novel contributions: 
\begin{enumerate} 
	\item[1)] A resilient tube-based MPC algorithm that integrates a control sequence buffer, specifically designed to address amplitude-unbounded FDI attacks on the S-C channel. Based on the definition of such FDI attacks we propose, this approach allows the system to maintain resilience even when attacks exceed the maximum amplitude constraints. 
	\item[2)] An integrated resilient mechanism combining attack detection, optimal control, and resource allocation. This mechanism effectively mitigates amplitude-unbounded FDI attacks while employing minimal computational resources, without sacrificing system optimality.
	\item[3)] An iterative computational approach that utilizes hypothesis testing and probabilistic modeling to determine the maximal duration of an over-threshold FDI attack, assisting in selecting optimal buffer lengths and giving the theoretical feasibility guarantee of the resilient control strategy.
\end{enumerate}

	\section{Problem formulation and preliminaries}

\subsection{Notations}
The following notations are used throughout this paper. Define set addition as $A \oplus B \triangleq \{ a + b \,|\, a \in A ,\, b \in B  \}$ and set subtraction as $A \ominus B \triangleq \{ a \,|\, a  + B \subseteq A \}$.
Define the distance between point $x$ and set $Y$ as $d(x,\, Y) \triangleq \inf \{ \| x - y \|  \quad | y \in Y \} $.
Consider a discrete-time system formulated as $x^+ = f(x,w,\mathcal{a})$, the set $S$ is a robustly invariant set augmented for attacks if the successor state $x^+ \in S$ for all $x \in S,\, w \in W, \, \mathcal{a} \in \mathcal{A}$, where $W$ and $\mathcal{A}$ are the bounded compact sets for process noises and tolerable FDI attacks, respectively. The tolerable FDI attacks are those attacks that are within the input constraints of the controller. If the FDI attacks are out of the bounded compact set, the controller will only receive the upper bound as the input signal.
A set $\Theta$ is robustly exponentially stable for $x^+ = f (x, w, \mathcal{a}), w \in W, \mathcal{a} \in \mathcal{A}$, with a region of attraction $X_N$ if there exists a $c > 0$ and a $\epsilon \in (0,1)$ such that any solution $x(i)$ of $x^+ = f (x, w, \mathcal{a})$ with initial state $x(0) \in X_N$, admissible disturbance $w$ and attacks $\mathcal{a}$ within threshold ($w(i) \in W, \, \mathcal{a}(i) \in \mathcal{A}$  for all $ i \ge 0$) satisfies $d(x(i), \Theta) \le c\epsilon ^i d(x(0), \Theta)$ for all $i \ge 0$. The natural numbers from $a$ to $b$ are represented by $\mathcal{N}_{[a,b]}$ and the set of all natural numbers are denoted by $\mathcal{N}$.

\subsection{System dynamics}

In this paper, we investigate a CPS with uncertainties described using the discrete-time state space method. The sensor-to-controller channel of this system is susceptible to severe FDI attacks, which have unbounded amplitudes but are constrained by probability.

The system dynamics are given by the following equations\cite{Y.Liu2021}: 
\begin{equation}
x(k+1) = Ax(k) + Bu(k)+ F w(k),
\end{equation}
\begin{equation}
\tilde{x}(k) = \hat{\mathcal{f}}\big(x(k), \mathcal{a}^{SC}(k)\big),
\end{equation}
\begin{equation}
u(k) = \kappa(\tilde{x}(k)),
\end{equation}
where $x(k)  \in \mathbb{R}^n$ denotes the system state at time instant $k$ and $\tilde{x}(k) \in \mathbb{R}^n$ represents the tampered system state by malicious FDI attack signal $\mathcal{a}^{SC}(k) \in \mathbb{R}^n$ on the S-C channel. $u(k) \in \mathbb{R}^m$ is the control input signal applied to the system at time $k$, obtained through a certain control law $\kappa(\cdot)$. $w(k)  \in \mathbb{R}^d$ denotes the process disturbance at time instant $k$. The system matrices are defined as $A \in \mathbb{R}^{n \times n}$, $B \in \mathbb{R}^{n \times m}$, and $F \in \mathbb{R}^{n \times d}$, respectively.

The tampered system state signal is formulated as 
\begin{equation}\label{tampered state x}
\hat{\mathcal{f}}(\cdot) = \hat{\mathfrak{J}}(\mathfrak{a}(k),x(k)) + \mathfrak{a}(k) \mathcal{a}^{SC}(k),
\end{equation}
where $\mathcal{a}^{SC}(k)$ is the manipulation data injected into the S-C channel and the function $\hat{\mathfrak{J}}(\mathfrak{x},\mathfrak{y})\triangleq(1-\mathfrak{x})\mathfrak{y}$.
The random variable $\mathfrak{a}(k)$ obeys the Bernoulli distribution as $\mathfrak{a}(k)\sim B(1,\bar{\mathfrak{a}})$ with the mean value $\bar{\mathfrak{a}}$. It works as a sign and $\mathfrak{a}(k) = 1$ denotes that the data carrying true system state will be tampered with as the FDI signal $\mathcal{a}^{SC}(k)$.
In severe attack scenarios, the controller receives information with significant deviations, resulting in completely erroneous control decisions.

\begin{remark}
	This discrete-time, linear time-invariant system modeling is reasonable and efficient in the field of network security and attack defense \cite{Franze_severe}\cite{Y.Liu2021}. However, most CPSs in the real world exhibit varying levels of nonlinearity, and our theoretical analysis is based on linear systems, which inevitably results in limitations. Therefore, in practical applications, we need to linearize them around certain stable operating points based on the evaluation of actual systems. This linearization method has been validated in the test case in Section \ref{Nonlinear Case}.
\end{remark}

The term $w \in W$ represents bounded disturbances, where $W$ is a compact set containing the origin. The system state $x$ and actuator input $\tilde u$ are subject to the physical constraints $x \in \mathcal{X}$ and $\tilde u \in \mathcal{U}$, where $\mathcal{X}$ and $\mathcal{U}$ are also compact sets containing the origin.

\subsection{Mathematical model of FDI attack}\label{Mathematical model}

The cyber control system field is susceptible to frequent malicious cyber attacks. However, analyzing these detrimental incidents provides valuable insights for future defense strategies and enables the development of mathematical models for various cyber attacks. This paper focuses on discussing the model of FDI attacks based on probability theory.

In discrete-time control systems, FDI attacks manifest as probabilistic events where resource-constrained adversaries compromise subsystems intermittently. The number of successful attacks \(N_\mathcal{a}\) during \(n\) sampling intervals follows a binomial distribution:
\begin{equation}
	N_\mathcal{a} \sim B(n,\bar{\mathfrak{a}})
\end{equation}
where \(\bar{\mathfrak{a}}\) represents the probability of successful compromise per instant. This parameter inherently reflects three key characteristics of real-world attacks: attackers' limited resources lead to probabilistic target selection, components are compromised intermittently rather than continuously, and varying security levels across infrastructure result in different success probabilities. By modeling independent Bernoulli trials at each instant, this distribution establishes a unified analytical framework for attack dynamics in CPS security \cite{BinoD}\cite{BinoD2}\cite{Yu2024}.

The amplitude of FDI attacks on the S-C channel follows a Gaussian distribution:  
\begin{equation}
	\mathcal{a}_k \sim N(\mu, \sigma^2)
\end{equation}
where \(\mu\) denotes the mean and \(\sigma\) the standard deviation. To maximize stealth, we configure \(\mu = 0\) to mimic ambient noise characteristics, causing preliminary detectors to misclassify attacks as stochastic noise. The standard deviation \(\sigma\) exceeds historical noise variance by an order of magnitude (empirically >10×) to ensure attack effectiveness while respecting adversarial resource constraints. This parameterization captures the combined effect of attacker knowledge, resource limitations, and infrastructure vulnerabilities - phenomena whose aggregation satisfies Central Limit Theorem conditions for normality \cite{NormalD}. Section IV.C further validates parameter robustness through sensitivity analysis across \(\sigma \in [1,100]\), demonstrating consistent performance stability.

\begin{remark}
The binomial attack model finds concrete validation in real-world incidents such as the 2015 Ukraine grid cyberattack, where attackers propagated malware to compromise multiple substations \cite{Blazek2025}. In this physical scenario, the parameter $\bar{\mathfrak{a}}$ quantifies vulnerability exposure levels, representing the probability of successful compromise at each substation during the attack period. After gaining access, attackers injected normally distributed false data into SCADA systems, deliberately redirecting power flows to cause component overloads and cascading failures. This sequence ultimately triggered widespread blackouts affecting 230,000 residents \cite{Farahani2024}.
\end{remark}

\begin{proposition}
	For a CPS vulnerable to FDI attacks, there exists a threshold attack level, denoted as $\mathcal{a}_{th}$, below which the system will exhibit robust exponential stability with an invariant set $Z$.
\end{proposition}

\begin{remark}
This proposition enables us to address low-amplitude attacks using robust methods for bounded disturbance, thereby conserving resilient resources in common scenarios.
\end{remark}

By utilizing the properties of the Gaussian distribution, we can calculate the probability of events where the attack exceeds the threshold.

The probability that the attack’s amplitude exceeds the threshold, denoted as $\zeta$, is defined as:
\begin{equation}\label{overthreshold p}
    \zeta = P(|\mathcal{a}(t_k)| > \mathcal{a}_{th})= 2\int_{\mathcal{a}_{th}}^{+\infty} \frac{1}{\sqrt{2\pi}\sigma}e^{-\frac{(a-\mu)^2}{2\sigma^2}} da.
\end{equation}

Furthermore, the probability of FDI attacks occurring at time $t_k$ while simultaneously exceeding the threshold is given by $\mathfrak{a}\zeta$. Since events at different sampling instants are independent, we can determine the probability of attacks occurring at $j$ consecutive moments with an attack amplitude greater than the threshold, which is denoted as $(\mathfrak{a}\zeta)^j$.

\begin{remark}
It is important to note that the probability $(\mathfrak{a}\zeta)^j$ cannot be applied to the total of $N_{sim}$ sampling instants. The concept of “small probability events cannot occur” refers to events with a probability of less than 1\% that are unlikely to occur in a single experiment. However, with a sufficient number of independent repeated experiments, there is still a considerable probability of their occurrence. Further elaboration on this topic will be provided in Section \ref{probability theory}.
\end{remark}

\subsection{Tube-based MPC optimization problem}
The resilient control law employs a tube-based MPC framework with prediction horizon $N$ representing the optimization window length. The decision variable $(\overline{x}_k,\mathbf{u})$ consists of the initial state $\overline{x}_k$ of the optimal problem at time instant $k$  and the control sequence $\mathbf{u} := \{u_{k}, \dots, u_{k+N-1}\}$ driving the nominal system.

The framework selects optimal $\overline{x}_k^*$ within neighborhood $Z$ of actual state $x$, forming a "tube" to handle bounded disturbances. The optimization problem $\mathcal{P}_N^*(x)$ is:

\begin{align}
	\Upsilon_N^*(&x_k)=\operatorname*{min}_{\overline{x}_k,\boldsymbol{u}}:\Upsilon_N(\overline{\boldsymbol{x}}_{k},\boldsymbol{u})\\
	\mathrm{s.t.} \quad &\overline{x}_{k+i}\in (\mathcal{X}\ominus Z),\label{X-Z} \\ 
	&{ u_{k+i} }\in(\mathcal{U}\ominus K Z) \label{U-KZ},\\ 
	&\overline{x}_{k+i+1}=f(\overline{x}_{k+i}, u_{k+i}),i \in \mathcal{N}_{[0,N-1]},\label{x+ = f(x,u)} \\
	&\overline{x}_{k+N}\in X_f  \subset (\mathcal{X}\ominus Z),\label{Xf}\\ 
	&x_k\in (\overline{x}_k\oplus Z). \label{xk}
\end{align}

where $N$ is the prediction horizon length. $\Upsilon_N$ represents the value function in the optimization problem, $x_k$ is the measured system state obtained from sensors, and $\overline{x}_k$ denotes the states in the nominal system (hypothetical system in the controller).

\begin{remark}
Practical selection of $N$ should consider the minimum horizon ensuring recursive feasibility and the performance-computation trade off. Larger $N$ improves performance but increases computational burden, while smaller $N$ may violate recursive feasibility. For our case study (Section IV), $N=10$ was selected as it minimizes $\|x-x_{\text{ref}}\|$ while keeping solve time $< 0.3T_s$ ($T_s$: sampling period).
\end{remark}

For constraint settings, refer to\cite{Mayne2005}. The most intuitive explanation for state constraints (\ref{X-Z}) is that in order for the system state to still comply with the initial constraint conditions under bounded disturbances, a stricter constraint is required, subtracting a disturbance upper bound from the initial constraint conditions. Similarly, the effect of bounded disturbances on the constraints (\ref{U-KZ}) of control inputs requires subtracting a $KZ$, where $K$ is the state feedback matrix that ensures Lyapunov stability. (\ref{x+ = f(x,u)}) denotes the dynamic equation of nominal system based on system model. The terminal region $X_f$ in (\ref{Xf}) is set for this optimal problem to guarantee feasibility and stability. The selection of the optimal initial state in (\ref{xk}) constitutes a “tube”.

The cost function in this context is defined as:
\begin{align}
	&\Upsilon_N \triangleq\sum_{i=0}^{N-1}L(x_{k+i},u_{k+i})+F(x_{k+N}),\label{JN} \\
	&L(x_{k+i},u_{k+i}) \triangleq \frac{1}{2} (x^TQx + u^TRu), \label{L} \\
	&F(x_{k+N}) \triangleq  \frac{1}{2} x^TPx, \label{F}
\end{align}
where $Q$, $R$ and $P$ are positive definite matrices.

In our scheme, the aforementioned optimal control problem is solved online at each instant when an over-threshold FDI attack \textbf{does not} occur.

\begin{figure}[t]
	\centering
	\includegraphics[width=3.5in]{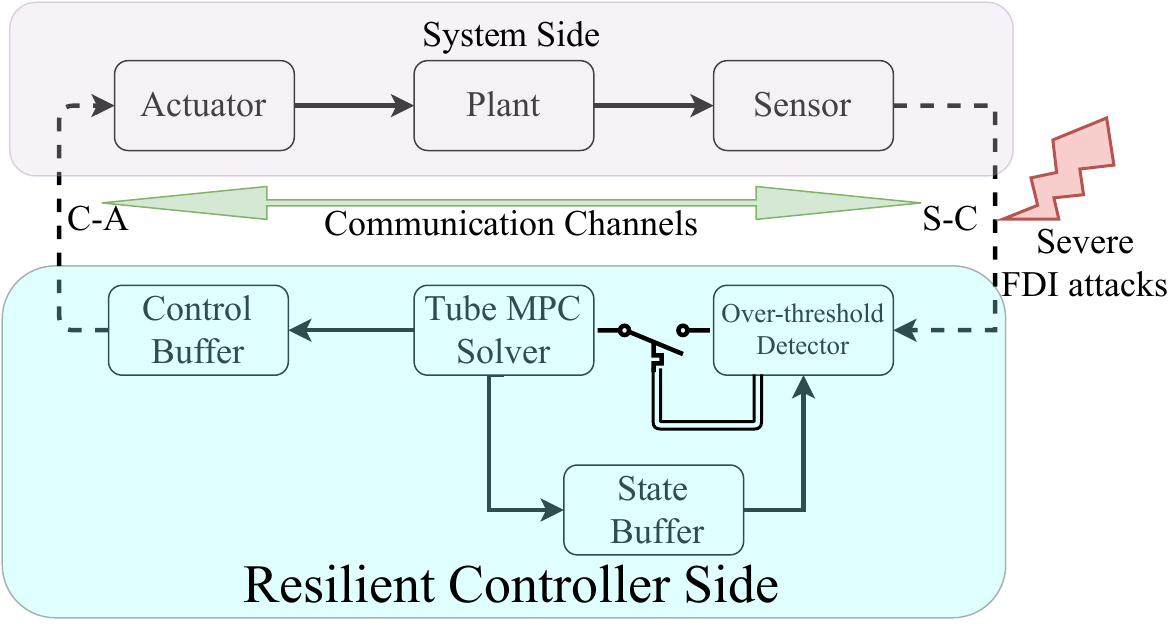}
	\caption{Resilient control architecture.}
	\label{architecture}
\end{figure} 

\subsection{Resilient control strategy}
We propose the following resilient control strategy for the scenario where the S-C channel of a CPS is subject to unbounded FDI attacks:
\begin{enumerate}
	\item During normal operation (no attack or when the attack is below the threshold), the \textbf{tube MPC solver} calculates optimal control and state sequences. These sequences are stored in the \textbf{control buffer} and \textbf{state buffer} for closed-loop control. The buffer lengths are determined by Section \ref{probability theory};
	\item The state sequences in the \textbf{state buffer} are fed back to the \textbf{over-threshold detector} for subsequent over-threshold attack detection. The control sequences in the \textbf{control buffer} are used for possible open-loop control in the future;
	\item  When the S-C channel is under an over-threshold FDI attack, the data input to the \textbf{tube MPC solver} is cut off, and the control sequences in the \textbf{control buffer} are used for open-loop control.
\end{enumerate}
﻿

The control strategy is specified in Fig. \ref{architecture}.
To illustrate the structure of the proposed resilient tube-based MPC scheme, we present it in Algorithm 1.

The control law of the proposed resilient stochastic tube-based MPC scheme can be formulated as follows:
\begin{equation}
	\begin{aligned}
		\kappa\big(x(t_k)\big) = &\hat{\mathfrak{J}}\Big(\nu(t_k), u\big(t_k|x(t_k)\big)\Big)\\ &+ \nu(t_k) \mathcal{u}\big(t_k|x(t_k-ct)\big),
	\end{aligned}
\end{equation}
where $\nu(t_k)$ is a random variable which obeys the Bernoulli distribution as $\nu(t_k) \sim B(1, \bar{\mathfrak{a}}\zeta)$. $u\big(t_k|x(t_k)\big)$ is the optimal control input based on current system state $x(t_k)$, while $\mathcal{u}\big(t_k|x(t_k-ct)\big)$ is the suboptimal control input based on $x(t_k-ct)$. The function $\hat{\mathfrak{J}}(\mathfrak{x},\mathfrak{y})\triangleq(1-\mathfrak{x})\mathfrak{y}$.

\begin{remark}
As an essential hyperparameter in Algorithm 1, the detection threshold $d_{th}$ depends on several factors. As seen in Algorithm 1, $d_{th} = \|W_{\mathcal{a}}\|(A_{th} + \tau \bar{w})$, where $\|W_{\mathcal{a}}\|$ is the weighing matrix of FDI attacks on the different states, $A_{th}$ is the attack threshold, $\tau$ is the coefficient of bound of disturbance $\bar{w}$. First, $A_{th}$ should approach the attack level $\|\mathcal{A}\|$. At least, it should not be less than the value of $\|\mathcal{A}\|$. Second, $\tau \bar{w}$ should describe the overall uncertainty of the system model, including process noises, measurement noises, model mismatch, and system nonlinearity as well. The greater the uncertainties, the larger coefficient $\tau$ should be taken. For example, in our nonlinear test case, larger $\tau$ and consequently larger $d_{th}$ are used compared to the linear case. Finally, the weighted matrix $\|W_{\mathcal{a}}\|$ is to weigh the impact of FDI attacks on each system state. This is decided by the attacker's resources and the system's nature.
\end{remark}

\begin{algorithm}[htbp]
	\renewcommand{\algorithmicrequire}{\textbf{Input:}}
	\renewcommand{\algorithmicensure}{\textbf{Output:}}
	\caption{Resilient MPC using attack detection}
	\label{power_compressed}
	\begin{algorithmic}[1]
		\REQUIRE $x(t_k)$, $\bar{\mathfrak{a}}$, $\zeta$, $A_{th}$, $W_{\mathcal{a}}$, $\tau$ and $\bar{w}$;
		\ENSURE control input $\kappa(x(t_k))$;
		\STATE Initialize $ct=1$, $\nu(t_k)=0$; Calculate $b$ and set $\lambda=b$;
		\STATE Calculate detection threshold $d_{th} = \|W_{\mathcal{a}}\|(A_{th} + \tau \bar{w})$;
		\WHILE {control does not stop}
		\STATE Obtain measured state $\tilde{x}(t_k)$; Set attack flag $\nu(t_k) = 0$;
		\IF {the \textbf{state buffer} is not empty}
		\STATE $d = \|\tilde{x}(t_k) - \mathcal{x}(t_k|t_k - ct)\|$;
		\IF{$d > d_{th}$} \STATE $\nu(t_k) = 1$; \COMMENT{Over-threshold attack detected}
		\ENDIF
		\ENDIF
		\IF {$\nu(t_k)=0$ \OR $ct > \lambda$} \STATE\COMMENT{Normal or Recovery Mode}
		\STATE Solve $\mathcal{P_N}(x(t_k))$ for optimal $\mathbf{u^*}(\cdot)$ and $\mathbf{x^*}(\cdot)$;
		\STATE Load first $\lambda$ elements to \textbf{control} and \textbf{state buffers};
		\STATE Apply control $\kappa(x(t_k))=u(t_k|x(t_k))$; Reset $ct = 1$;
		\ELSE \STATE\COMMENT{Resilient Mode}
		\STATE Apply buffered control $\kappa(x(t_k))=\mathcal{u}(t_k|x(t_k-ct))$;
		\STATE Increment $ct \gets ct+1$;
		\ENDIF
		\STATE Apply control $\kappa(x(t_k))$, increment $t_k \gets t_k+1$;
		\ENDWHILE
	\end{algorithmic}
\end{algorithm}

\subsection{Preliminary results}\label{Preliminary results}

Since our proposed scheme is a generalization of robust methods in FDI attack scenarios, it is necessary to review some relevant results from previous research \cite{Mayne2005} and \cite{MayneE2000}.
﻿
The preliminary axioms of stability analysis (\ref{pre1}), (\ref{pre2}) and (\ref{pre3}) are utilized outlined below:
\begin{equation}\label{pre1}
	\Upsilon^*(x) \geq \sigma_1\|x_0^*(x)\|^2, \quad \forall x \in X_N.
\end{equation}
\begin{equation}\label{pre2}
	\begin{aligned}
		&\Upsilon^*(x^+)-\Upsilon^*(x) \leq -\sigma_1\|x_0^*(x)\|^2, \\
		&\forall x \in X_N, \forall x^+ \in (Ax + B\kappa^*(x)) \oplus W .
	\end{aligned}
\end{equation}
\begin{equation}\label{pre3}
	\Upsilon^*(x) \leq \sigma_2 \|x_0^*(x)\|^2, \quad \forall x \in X_f \oplus Z.
\end{equation}
\begin{remark}
	These axioms delineate the properties of the upper and lower bounds for the optimal value function, thereby ensuring the absolute decrease in the value of the optimal function during the rolling optimization process. They offer foundational prerequisites for the stability of tube-based MPC in non-attack scenarios.
\end{remark}

\begin{theorem}
	For CPS with bounded uncertainty and no attacks, the set $Z$ around the origin exhibits robust asymptotic stability, with the feasible domain $X_N$ serving as the attractive region.
\end{theorem}

\begin{IEEEproof}
	Define $\varrho \triangleq 1-\sigma_1 / \sigma_2 \in (0,1)$ where $\sigma_2 > \sigma_1$. The solution of $x^+= Ax+B \kappa^*(x)+w$ yields $x(i)$ for all $i \in \mathcal{N}$. Then define a scalable set $\Omega_a \triangleq \{ x | \, \Upsilon^*(x)  \leq a, \, \forall a > 0 \}.$ Based on the definition of $Z$, it follows that when $a=0$, the set $\Omega_a= \Omega_0=Z$. By gradually increasing $a$, the set $\Omega_a$ expands accordingly. We can always find an $a$ such that $\Omega_a \subset Z \oplus X_f$, where inequality (\ref{pre3}) holds. It is evident that within this region, $x \in X_N$ always holds because $X_f \oplus Z \subset X_N$, satisfying the conditions of (\ref{pre1}) and (\ref{pre2}).
	
	From equations (\ref{pre2}) and (\ref{pre3}), we obtain $\Upsilon^*(x(k+1)) \leq (1-\frac{\sigma_1}{\sigma_2})\Upsilon^*(x(k))=\varrho \Upsilon^*(x(k))$ where $k \in \mathcal{N}$ and $x(0) \in \Omega_a$. From recursion, we can infer $\Upsilon^*(x(i)) \leq \varrho^i \Upsilon^*(x(0)), \varrho \in (0,1)$. Next, using inequalities (\ref{pre1}) and (\ref{pre3}), we can derive $\|x^*_0(x(i))\| \leq c\sqrt{\varrho}^i\|x^*_0(x(0))\|,\,\forall x(0) \in \Omega_a$ for some constant $c<\infty$ and $\sqrt{\varrho} \in (0,1), i\in \mathcal{N}$. We can always find a constant $\mathcal{I}$ such that for all $i \geq \mathcal{I}$, $x(i) \in \Omega_a \subset X_N$ holds. Hence, there exists a greater finite constant $c_1>c$ such that $\|x^*_0(x(i))\| \leq c_1\sqrt{\varrho}^i\|x^*_0(x(0))\|,\,\forall x(0) \in X_N$. According to the definition, the theorem is valid.
\end{IEEEproof}

Theorem 1 verifies that the system is initially robustly asymptotically stable without the impact of FDI attacks. This is a prerequisite for conducting our resilient control scheme.
	\section{Theoretical results}
\subsection{Buffer length design based on probability theory}\label{probability theory}
In this subsection, we first propose a probabilistic problem below. Then we solve it to obtain the parameter to design the length of our \textbf{control buffer}.

\textbf{Probabilistic Problem}:
Given the probability of an over-threshold attack occurring at a single sample instant as $\bar{\mathfrak{a}}\zeta$, determine the maximum number of consecutive occurrences of such attacks, denoted as $b$, within a finite time horizon $N$, with a significance level of $\alpha$.

\textbf{Event $A$}:
Consecutive over-threshold FDI attacks occur $b$ times in a total of $N$ instants.
﻿
Define the probability of event $A$ as $P_N^b$:
\begin{equation}
	P(A) \triangleq P_N^b .
\end{equation}

To solve the Probabilistic Problem, we can divide event $A$ into several sub-events based on the starting instant of the consecutive over-threshold attacks. These consecutive attacks may occur at instants $1,2,\cdots, N-b+1$. Therefore, we define the sub-event:

\textbf{Sub-event $A_k$}:
Event $A$ with a starting instant of $k$, where $k \in \mathcal{N}_{[1,N-b-1]}$.

The relationship between $P(A)$ and $P(A_k)$ is as follows:
\begin{equation}
	P(A) = \sum_{k=1}^{N-b-1}P(A_k).
\end{equation}

It is evident that once the over-threshold attacks occur consecutively $b$ times starting at instant $k$, we no longer need to consider future events since event $A$ has already occurred. The only relevant factor is the situation before event $A$. Hence, the probability of the sub-event can be represented as:
\begin{equation}
	P(A_k)=
	\begin{cases}
		(\bar{\mathfrak{a}}\zeta)^b, &k=1\\
		(\bar{\mathfrak{a}}\zeta)^b(1-\bar{\mathfrak{a}}\zeta)(1-P_{k-2}^b), &k\in\mathcal{N}_{[2,N-b+1]}
	\end{cases}.
\end{equation}
where $\bar{\mathfrak{a}}$ is the mean value of random variable $\mathfrak{a}$ conforming to Bernoulli distribution defined in (\ref{tampered state x}). and $\zeta$ is the probability that the unbounded FDI attack's amplitude exceeds the threshold $\mathcal{a}_{th}$, defined in (\ref{overthreshold p}).
$P_{k-2}^b$ represents the probability of the event “consecutive over-threshold FDI attacks occurring $b$ times in a total of $(k-2)$ time instants” (if $k\leq 2$, then $P_{k-2}^b = 0$). It is evident that this is a recursive problem in calculating $P(A)$, as shown below:
﻿
\begin{equation}\label{概率递推公式}
	P(A)=P_N^b=(\bar{\mathfrak{a}}\zeta)^b\Big(1 + \sum_{k=2}^{N-b+1} (1-\bar{\mathfrak{a}}\zeta)(1-P_{k-2}^b) \Big).
\end{equation}

According to the principle of small probabilities, when $P_N^b<\alpha$ (with a significance level of $\alpha$ chosen as 1\% in this paper), the system cannot be subjected to attacks greater than the threshold for $b$ consecutive moments.

Hence, we can set the length of the proposed \textbf{control buffer} as 
\begin{equation}\label{find lambda}
	\lambda = \min \{b \mid P^b_N < \alpha\}.
\end{equation}

\begin{remark}
	When the system is subjected to $b$ consecutive over-threshold FDI attacks, we switch the system to open-loop mode for $b$ successive sampling instants and use the $\lambda$ signals in the \textbf{control buffer} to drive the actuator. Thus, we can assert that the control scheme is always feasible under such modeled FDI attacks because the control input sequence $\mathcal{u}$ stored in the buffer will never be depleted even with a maximum of $b$ consecutive over-threshold FDI attacks.
\end{remark}

\subsection{Input-to-state stability analysis for FDIs under threshold}

In this subsection, we consider the scenario where only FDI attacks that are within the threshold will occur. Based on this scenario, we discuss the input-to-state stability (ISS) of the set $Z$.

To establish the validity of the system's ISS, we need to make certain assumptions regarding the stage cost and terminal cost of the value function \cite{Sunqi_phd}. This is a common practice in most MPC studies to ensure ISS and robustness.

\begin{assumption}
	\begin{equation}\label{Assumption for ISpS}
		\mathcal{V}_f(f(x,\kappa^*(x)+\mathcal{a},w)) - \mathcal{V}_f(x) \leq -\mathcal{l}(x,\kappa^*(x))+ \alpha\|\mathcal{A}\|,
	\end{equation}
	where $\| \mathcal{A}\| \triangleq sup_{\mathcal{a} \in \mathcal{A}}\|\mathcal{a}\|$ represents the supremum of the within-threshold FDI attacks and $\alpha$ is a positive constant. 
\end{assumption}

In other words, $\| \mathcal{A}\|$ can be interpreted as the threshold of FDI attacks considered in this subsection.

\begin{remark}
	In this constrained attack scenario, $\|\mathcal{A}\| = A_{th}$. This assumption is made based on the decrement property of the terminal cost. When the FDI attack is canceled ($\|\mathcal{A}\|=0$), it aligns with the conventional axiom observed in most MPC studies. The conventional axiom ensures that the terminal cost function strictly decreases by the amount of the single-step stage cost. However, when the system is exposed to an attack, this property may be diminished by an amount related to the supremum of $\mathcal{a}$. Thus, to guarantee the validity of this assumption, the energy level of the FDI attack should be limited, which precisely aligns with the scenario considered in this subsection.
\end{remark}

\begin{assumption}
\begin{equation}\label{lower bound of l}
	\mathcal{l}(x,\kappa^*(x)) \geq \beta \|x_0^*(x)\|^2.
\end{equation}
where $\mathcal{l}$ represents the stage cost and $\beta$ is a positive constant.
\end{assumption}

\begin{remark}
	This assumption defines the infimum of the stage cost function as the norm of the optimal initial state $x_0^*(x)$. This is crucial as it enhances the diminishing property of the optimal value function $\Upsilon^*_N(x)$. Without this assumption, the optimal value function may lose its decrement property in the presence of an FDI attack.
\end{remark}

Then we propose the following theorem.

\begin{theorem}
	For bounded uncertain CPSs exposed to FDI attacks within the threshold, if Assumptions 1 and 2 hold, the set $Z$ remains robustly ISS, with the attractive domain being the feasible domain $X_N$.
\end{theorem}

\begin{IEEEproof}
	Using (\ref{Assumption for ISpS}) in the $N-1$ predicted step, we have $		\mathcal{l}(x(N-1),\kappa^*(x(N-1))) +  \mathcal{V}_f(x(N)) -  \mathcal{V}_f(x(N-1)) \leq \alpha\|\mathcal{A}\|$. Rewriting the stage cost and combining like terms, we obtain $	\Upsilon^*_N(x) - \Upsilon^*_{N-1}(x) \leq \alpha\|\mathcal{A}\|,\quad \forall x \in X_N$.
	Consequently, we can derive the decrement property of the optimal value function between real-time instants (from $x$ to $x^+$). Since $\Upsilon^*_{N}(x) = \Upsilon^*_{N-1}(x^+) + \mathcal{l}(x,\kappa^*(x))$ holds, we can obtain $\Upsilon^*_N(x^+) - \Upsilon^*_{N}(x) \leq \alpha\|\mathcal{A}\| + \mathcal{l}(x,\kappa^*(x))$. (\ref{lower bound of l}) allows us to prove that $\Upsilon^*_N(x^+) - \Upsilon^*_{N}(x) \leq -\beta\|x_0^*(x)\|^2 +\alpha\|\mathcal{A}\|.$ Then we can iteratively obtain $\Upsilon^*(x(i)) \leq \rho^i\Upsilon^*(x(0)) + \psi\|\mathcal{A}\|$, where  $\rho = 1-{\beta}/{\sigma_2}$ and $\psi = {(1-\rho^i)\alpha}/(1-\rho)$, $i \in \mathcal{N}$.
	Finally, we can easily find constants $c_1,c_2 \in (0,\infty)$ satisfying
	$
	\|x^*_0(x(i))\| \leq c_1\sqrt{\rho}^i\|x^*_0(x(0)) \|+ c_2
	\|\mathcal{A}\|
	$.
	The theorem has been proven through the definition of the ISS of the set $Z$.
\end{IEEEproof}

Theorem 2 verifies that under the scenario of bounded FDI attacks if the attacked system meets the necessary assumptions, the stability metric of the CPS transitions from robust asymptotic stability (in the absence of FDI attacks) to robust ISS (under bounded FDI attacks). Additionally, the domain of attraction remains the feasible region for the optimal problem $\mathcal{P}_N^*(x)$.

\subsection{Feasibility and stability analysis for over-threshold FDIs}

In the previous subsection, we established the stability of the system when it experiences FDI attacks below the threshold. In this case, we can identify a feasible region $X_N$ for the initial states, ensuring that there exists a control sequence $\mathbf{u}$ that satisfies the control input constraint. To distinguish it from the newly proposed feasible region in this section, we refer to this region as $X_N^0$. For all states $x$ in $X_N^0$, any sequence $\mathbf{u}$ can form an admissible control input set $\mathcal{U}_N^0$. To clarify:
\begin{equation}
    X_N^0 \triangleq \{x|\, U_N^0 \neq \emptyset \},
\end{equation}
\begin{equation}
    \mathcal{U}_N^0 = \{ \mathbf{u} | \, u(i) \in \mathcal{U},\, x^*(i,\mathbf{u}) \in \mathcal{X} ,\, x^*(N,\mathbf{u}) \in \mathcal{X}_f  \},
\end{equation}
where $i \in \mathcal{N}_{[0,N-1]}$.

It is important to note that $X_N^0$ represents the feasible region when all FDI attacks are under the preset threshold and is not applicable in cases where random over-threshold attacks occur. Hence, we introduce a new region denoted as $X_N^\lambda$ to represent the admissible initial state set in the presence of at most $\lambda$ consecutive over-threshold FDI attacks.

From Algorithm 1, we can identify a control sequence $\mathbf{u} = \{ u(1), u(2),\cdots u(\lambda),u^*(\lambda+1),\cdots u^*(N) \}$. This sequence contains the first $\lambda$ feasible control inputs solved at historical instants and $N-\lambda$ optimal control inputs solved at the present. Both the feasible and optimal control inputs satisfy the constraints.

Based on this control sequence, we define a new control input set as follows: 
\begin{equation}\label{可行控制集的定义}
\begin{aligned}
         \mathcal{U}_N^\lambda = \{ \mathbf{u} | \, &u(i) \in \mathcal{u},\forall i \in [1,\lambda], \\
      &u(i) \in \mathcal{U},\forall i \in [\lambda+1,N], \\
      &x^*(i,\mathbf{u}) \in \mathcal{X},\forall i \in [0,N-1],\\
      &x^*(N,\mathbf{u}) \in \mathcal{X}_f  \},
\end{aligned}
\end{equation}
where $\mathcal{u}$ represents the control input sequence stored in the control buffer. This means that even in the worst-case scenario of consecutive $\lambda$ over-threshold attacks, we can still find a feasible control input sequence to guide the system’s state from the initial region to the terminal set.

\begin{remark}
	In the work of \cite{Sunqi_phd} addressing DoS attacks, the composition of the allowable control set is filled by zero control inputs, reflecting the absence of communication. In contrast, when addressing over-threshold FDI attacks in our framework, the admissible control set comprises up to $\lambda$ feasible control inputs, alongside $N-\lambda$ optimal control inputs. This highlights the robustness of our approach when compared to previous work on DoS attacks.
\end{remark}

The corresponding feasible state set is defined as:
\begin{equation}\label{definition of X_N^LAM}
    X_N^\lambda \triangleq \{x| \, \mathcal{U}_N^\lambda \neq \emptyset\}.
\end{equation}

\begin{assumption}
	Considering the FDI attack scenario modeled in Section \ref{Mathematical model}, the initial feasible region $X_N^\lambda$ is not empty for each calculated $\lambda$.
\end{assumption}

\begin{remark}
	This assumption ensures that the control problem is feasible even in the worst-case scenario. The reason why this situation is considered the worst-case is that the maximum number of consecutive occurrences ($\lambda$) happens at the beginning of the horizon $N$ when the system is farthest from the equilibrium steady state.
\end{remark}

Next, we introduce the following lemma indicating recursive feasibility:
\begin{lemma}
If Assumption 3 holds, then the following recursive set dependency holds:
\begin{equation}\label{X_N recursive set dependency}
    X_N^\lambda \subseteq X_N^{\lambda-1} \subseteq \cdots \subseteq X_N^0 = X_N.
\end{equation}
\end{lemma}

\begin{IEEEproof}
	Referring to the definition of $\mathcal{U}_N^\lambda$ in (\ref{可行控制集的定义}), we observe that the only difference between $\mathcal{U}_N^\lambda$ and $\mathcal{U}_N^{\lambda-1}$ is the $\lambda$th term. Note that if $u(\lambda) \in \mathcal{u}$, then it must also belong to $\mathcal{U}$. Hence, we have $\mathcal{U}_N^\lambda \subseteq \mathcal{U}_N^{\lambda-1}\subseteq \cdots \subseteq \mathcal{U}_N^0 = \mathcal{U}_N$. From equation (\ref{definition of X_N^LAM}), we can conclude that the recursive set dependency in equation (\ref{X_N recursive set dependency}) is valid. The recursive feasibility is proven.
\end{IEEEproof}

From this point forward, we investigate the ISS of the resilient MPC scheme under over-threshold FDI attacks. The main difference in the ISS analysis compared to the previous subsection is the contraction of the feasible region.

\begin{theorem}
	For bounded uncertain CPSs exposed to FDI attacks modeled in Section \ref{Mathematical model}, assuming all the aforementioned assumptions hold, the system is ISS under the resilient tube-based MPC scheme, and the region of attraction is $X_N^\lambda$.
\end{theorem}

\begin{IEEEproof}
Firstly, we should ensure the feasibility of the resilient tube-based MPC scheme. As proven in Theorem 2, we establish feasibility by letting the initial state $x_0$ lie in $X_N$ when the attack is within the threshold. Similarly, we can extend this conclusion to $x_0 \in X_N^\lambda$ when over-threshold attacks may occur.

Based on the conclusions from the previous subsection, we can directly present the following useful inequality:
\begin{equation}
      \Upsilon^*(x) - \Upsilon^*(x^+) \geq \beta\|x_0^*(x)\|^2 - \alpha\|\mathcal{A} \|.
\end{equation}

This inequality illustrates that despite the occurrence of over-threshold FDI attacks in the S-C channel, the optimal value function maintains its monotonic decreasing property, with a minimum decrement of $\beta\|x_0^*(x)\|^2 - \alpha\|\mathcal{A} \|$.

Next, we can prove the exponential stability of $Z$ by contradiction. Assuming that, for an initial state lying in $X_N^\lambda$, it will not enter $\mathcal{X}_f \oplus Z$ in finite instants, we can find a $\bar{k} \in (0,\infty)$ such that $\Upsilon^*(x_0) < \bar{k}(\beta\|x_0^*(x)\|^2 - \alpha\|\mathcal{A}\|)$. Then, when $k>\bar{k}$, we can observe that the optimal function will decrease more than $k(\beta\|x_0^*(x)\|^2 - \alpha\|\mathcal{A}\|)$ and become less than 0, which contradicts its non-negativity. Hence, the subsequent proof of the ISS of the set $Z$ follows a similar approach to that in the previous subsection.
\end{IEEEproof}

\subsection{Terminal Conditions}
This subsection considers the construction of the terminal region and the associated terminal cost matrix of our resilient MPC framework.
We will employ linear matrix inequalities (LMIs), which are mathematically equivalent to Assumption 1, based on the properties of the Schur complement discussed in \cite{LMI}. The use of LMIs offers the advantage of simplifying the optimization problem associated with obtaining the terminal cost function involved in Algorithm 2.
\begin{lemma}
	The inequality in Assumption 1
	$$	\mathcal{V}_f(f(x,\kappa^*(x)+\mathcal{a},w)) - \mathcal{V}_f(x) \leq -\mathcal{l}(x,\kappa^*(x))+ \alpha\|\mathcal{A}\| $$
	holds $\forall w \in W$, for positive definite matrix $P$ and for some constant $\alpha\|\mathcal{A}\|$
	if the following LMI holds:
\begin{equation}\label{LMI}
	\resizebox{0.4\textwidth}{!}{$
		\begin{bmatrix}
			P^{-1} & 0 & (AP^{-1}+BKP^{-1})^\top & P^{-1} & (KP^{-1})^\top \\
			0 & \alpha\|\mathcal{A}\| & w^\top & 0 & 0 \\
			* & * & P^{-1} & 0 & 0 \\
			* & * & * & Q^{-1} & 0 \\
			* & * & * & * & R^{-1}
		\end{bmatrix}
	$}	\succeq 0.
\end{equation}

	\label{LMI assumption}
\end{lemma}

\begin{IEEEproof}
	This follows from the application of the Schur complement to (\ref{Assumption for ISpS}). Specifically, we can conclude that $\mathcal{Q}-\mathcal{S}\mathcal{R}^{-1} \mathcal{S}^\top\geq0$, which is equivalent to (\ref{LMI}). Here, $\mathcal{Q} = \text{diag}(P^{-1},\,\alpha\|\mathcal{A}\|)
$, $\mathcal{S} = \left[ \begin{array}{ccc} (AP^{-1}+BKP^{-1})^\top&P^{-1}&(KP^{-1})^\top\\
	w^\top&0&0
\end{array} \right]$ and $\mathcal{R}^{-1} =  \text{diag}(P,\, Q,\, R) $.
\end{IEEEproof}

\begin{lemma}\label{PIS}
	If set $\mathcal{S}_0 \triangleq \{ x|x \in \mathcal{X}\ominus Z, Kx \in  \mathcal{U}\ominus KZ\}$ and
	$\mathcal{S}_k  \triangleq \{ x | (A+BK)^i x \in \mathcal{S}_0 , i = 1,2,\cdots, k\}
	$ exist.
	It follows by 
	$\mathcal{S}_{k+1} = \mathcal{S}_k \cap \{ x|  (A+BK)^{k+1} x \in \mathcal{S}_0\} 
	$ and
	$\mathcal{S}_{k+1} \subseteq \mathcal{S}_{k}, k=0,1,2 \cdots.
	$
	Then there exists a finite constant $\xi$ that makes the equation $\mathcal{S}_{\xi+1} = \mathcal{S}_{\xi}$ hold.
\end{lemma}

\begin{IEEEproof}
	Suppose $\mathcal{S}_0$ is bounded, then $\mathcal{S}_k(k=0,1,2\cdots)$ is bounded since 	$\mathcal{S}_{k+1} \subseteq \mathcal{S}_{k}, k=0,1,2 \cdots
	$. If $\mathcal{X}\ominus Z$ and $\mathcal{U}\ominus KZ$ are not empty, $\mathcal{S}_0$ has the origin in its interior and is close set. This follows that $\mathcal{S}_0$ is compact and so is $\mathcal{S}_k$. There exists $r_1>0$, we have $\|x\| \leq r_1,  \forall x \in \mathcal{S}_k$. Since $(A+BK)$ is stable, for all $\epsilon >0$, no matter how small it is, there exists a $k$ (large enough) such that $\|(A+BK)^{k+1}x\| \leq \|A+BK\|^{k+1}\|x\|\leq \epsilon r_1$. 
	Since $\mathcal{S}_0$ is compact, we can find another constant $r_2>0$, such that a sphere with radius $r_2$ is in $\mathcal{S}_0$, i.e.$\{x|\|x\| \leq r_2 \} \subseteq  \mathcal{S}_0.$ We can always find an $\epsilon$ such that $\|(A+BK)^{\xi+1}x\| \leq \epsilon r_1 \leq r_2 , \forall x \in \mathcal{S}_\xi$, yielding $(A+BK)^{\xi+1}x \in \{x|\|x\| \leq r_2 \} \subseteq \mathcal{S}_0$, i.e. $\mathcal{S}_\xi \subseteq \mathcal{S}_{\xi+1}$. And since $\mathcal{S}_{\xi+1} \subseteq \mathcal{S}_\xi$, $\mathcal{S}_{\xi+1} = \mathcal{S}_{\xi}$ is valid.
\end{IEEEproof}

\begin{theorem}
	The terminal set $X_f$ is positively invariant and can be obtained if $P$ satisfies Lemma \ref{LMI assumption} and Lemma \ref{PIS} holds.
\end{theorem}

\begin{IEEEproof}
	LMI (\ref{LMI}) implies that $\max_{x\in X_f} \|x\|^2_{Q+K^\top RK}\geq \alpha\|\mathcal{A}\|$, yielding $\|x(k+1)\|^2_P \leq \|x(k)\|^2_P$. Then according to Lemma \ref{PIS}, we can choose $\mathcal{S}_{\xi}$ as a candidate of $X_f$.
\end{IEEEproof}

We can calculate matrix $P$ and terminal region $X_f$ by the procedure summarized in Algorithm \ref{Calculation of P Xf}.

\begin{algorithm}[htbp]
	\renewcommand{\algorithmicrequire}{\textbf{Input:}}
	\renewcommand{\algorithmicensure}{\textbf{Output:}}
	\caption{Computation of terminal region and cost matrix}
	\label{Calculation of P Xf}
	\begin{algorithmic}[1] 
		\REQUIRE  The parameters defining the system model $A$, $B$, $W$, attack level $\|\mathcal{A}\|$, the cost matrices $Q$, $R$; 
		\ENSURE $K$, $P$, $X_f$ ; 
		\STATE Solve OP $(K^{*},P^{*},\alpha^{*})=\operatorname*{min}_{S,Y,\alpha}\alpha\|\mathcal{A}\|\mathrm{~s.t. LMI (\ref{LMI})}$; 
		\STATE Initialize index $k = 0$, label $\nu = 0$;
		\WHILE {$\nu$ = 0}
		\IF {$(A+BK)^{k+1}x \in \mathcal{S}_0 , \forall x \in \mathcal{S}_k$}
		\STATE Set $\nu = 1$ to exit loop;
		\ELSE
		\STATE $k=k+1$;
		\ENDIF
		\ENDWHILE
		\STATE Set $X_f = \mathcal{S}_k$.
	\end{algorithmic}
\end{algorithm}

	\section{Illustrative example}

We conducted a simulation on a discrete-time harmonic oscillator system controlled by the proposed resilient tube-based MPC controller through a communication channel exposed to random amplitude-unbounded FDI attacks. The experiment was performed on Windows 10 using an Intel Core™ i7-9750H processor with MATLAB R2021a.

\subsection{System model and constraints}\label{System model}
Considering a mass-spring-damping system, described by:
\begin{equation}
	m\ddot{x}+ \mathfrak{F}_1(\dot{x}) + \mathfrak{R}_2(x) = u(t),
\end{equation}
where $x$ is the displacement of mass $m$; $\mathfrak{F}_1(\dot{x}) = c \dot{x}$ is the friction force ($c >0$); and $\mathfrak{R}_2(x) = kx + ka^2x^3$ is the spring's restoring force ($k, a >0$). With state $x(t) = [ x \ \dot{x} ]^\top$, the state-space model is
$$
\dot{x}(t) = \begin{bmatrix}
		0 & 1 \\
		-\frac{ \partial\mathfrak{R}(x) }{\partial mx } & -\frac{ \partial\mathfrak{F}(\dot{x}) }{\partial m\dot{x} }
	\end{bmatrix}
	x(t) + \begin{bmatrix}   0\\\frac{1}{m}	\end{bmatrix}
	u(t) +w(t),
$$
where $u(t)$ is the control input and $w(t)$ is a bounded disturbance. The state, control, and disturbance are constrained as follows:
\begin{align}
	x &\in \mathcal{X} \triangleq \{x| [I \ -I]^\top x \leq [\overline{x}^\top \ -\underline{x}^\top]^\top \}, \label{eq:X_Constraint} \\
	u &\in \mathcal{U} \triangleq \{u| [I \ -I]^\top u \leq [\overline{u}^\top \ -\underline{u}^\top]^\top \}, \label{eq:U_Constraint} \\
	w &\in W \triangleq \{w| [I \ -I]^\top w \leq [\overline{w}^\top \ -\underline{w}^\top]^\top \}. \label{eq:W_Constraint}
\end{align}
Based on Section III-A, we set the prediction horizon $N_p = 10$. Other parameters are in Table \ref{tab:hyperparameters}. 
This model is relevant for autonomous driving and smart building design \cite{M.Li2019}.

\begin{table}[H]
	\centering
	\caption{List of Hyperparameters}
	\label{tab:hyperparameters}
	\setlength{\tabcolsep}{4pt} 
	\begin{tabular}{lclc}
		\toprule
		\textbf{Parameter} & \textbf{Value} & \textbf{Parameter} & \textbf{Value} \\
		\midrule
		State bounds $\overline{x}, -\underline{x}$  &  $[5, 5]^\top$ & Mass $m$ & $1 \, \text{kg}$ \\
		Control bound $\overline{u}, -\underline{u}$ &  $2 \, \text{N}$ & Friction $c$ & $1.6 \, \text{N}\cdot\text{s/m}$ \\
		Disturbance bound $\overline{w}, -\underline{w}$ & $[0.05, 0.05]^\top$ & Spring const. $k$ & $1 \, \text{N/m}$ \\
		Hardening const. $a$ & $0.2 \, \text{m}^{-1}$ & Sampling time $T_s$ & $100 \, \text{ms}$ \\
		Initial state $x_0$ & $[2, -3]^\top$ & Horizon $N_p$ & $10$ \\
		\bottomrule
	\end{tabular}
\end{table}

\subsection{Attack pattern recognition and buffer length setting}

In this case, we modeled the FDI attacks through past data and analyzed their statistical distribution. Based on Section \ref{probability theory}, we determine that the probability of a successful triggered attack is $\bar{\mathfrak{a}}=0.2$, and the amplitude of the attack follows a normal distribution, $\mathcal{a}_k \sim N(0, 20^2)$. We choose a severe threshold $A_{th} = \sigma/5 = 4$, indicating that the probability of an attack exceeding the threshold is approximately $16.83\%$. To ensure conservatism, we set the significance level $\alpha = 0.01$. The total simulation step $N_{sim} = 100$. Using equation (\ref{概率递推公式}), we determine that the duration for which consecutive attacks are **statistically improbable** (at the given significance level) is $b \ge 5$.

\subsection{Numerical results}
\begin{figure}[htbp!]
	\centering
	\includegraphics[width=3in]{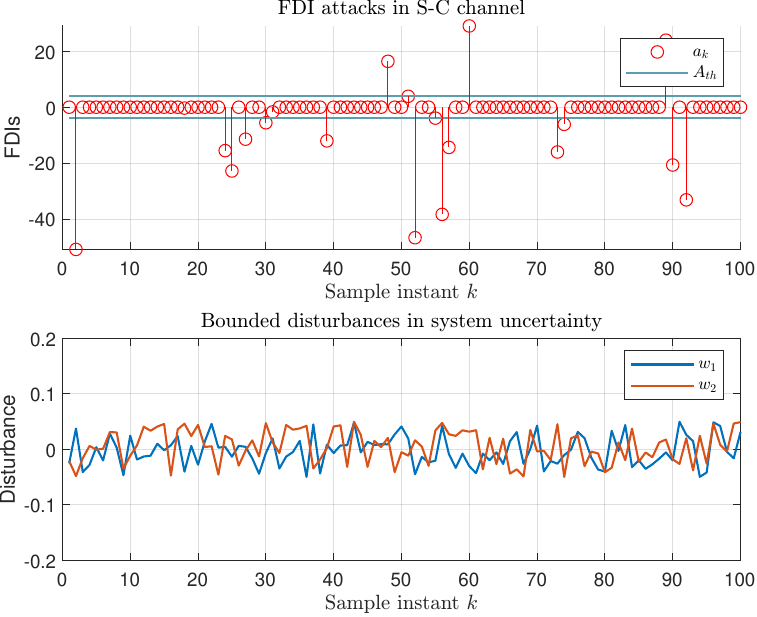}
	\caption{FDI attacks (unbounded amplitude); detection threshold $A_{th} = 4$; bounded process disturbances $\bar{w} = 0.05$.}
	\label{FDI&w}
\end{figure}

Through the simulation, Fig. \ref{FDI&w} shows the false data attacks injected in the S-C channel, along with the disturbances caused by system uncertainty.

The over-threshold attack detection method can identify whether the system is under FDI attacks that exceed the threshold in real time. We conduct simulations to evaluate the performance of the proposed method.

\begin{figure}[htbp!]
	\centering
	\includegraphics[width=3.5in]{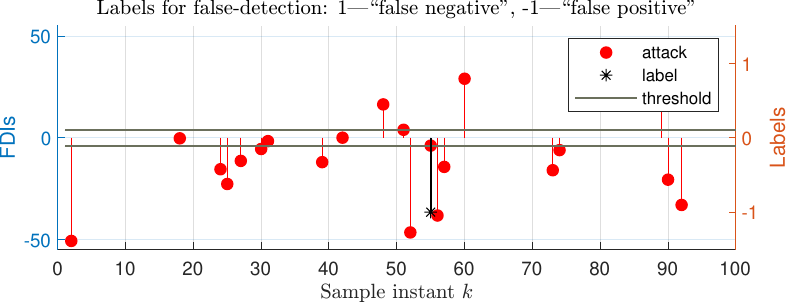}
	\caption{Over-threshold detection for FDI attacks. Over-threshold detection for FDI attacks. ``false negative" indicates that attacks occurred but were not recognized. ``false positive" indicates mistaking normal signals for anomalies.}
	\label{Detect}
\end{figure} 

To evaluate the accuracy of the proposed attack detector, we compare the number of detector alarms with the actual cases. The comparison threshold $d_{th} = \|W_{\mathcal{a}}\|(A_{th} + \tau \bar{w})$ is used to compare the distance between $\Tilde{x}$ and $\mathcal{x}$ in the buffer. It is important to note that $\tau$ is the adjustable parameter to balance conservatism. In this case, we choose $\tau = 2$ so that $d_{th} =  5.7983$.

We take 100 times Monte Carlo experiments to validate the performance of the proposed attack over-threshold detection mechanism. Most simulations over $N_{sim} = 100$ show no false detection. We select the worst case to show our analysis in Fig. \ref{Detect}. 
The results over $N_{sim} = 100$ show no cases of “false negative”, indicating that attacks occurred but were not recognized, and 1 instance of ``false positive", indicating mistaking normal signals for anomalies. Hence, the overall accuracy of the detection is 99\% in this scenario. It is worth mentioning that “false positives” may lead to poor performance resulting from executing feasible control inputs stored in the buffer instead of the optimal ones. “false negatives” will directly expose the system to attacks, causing serious consequences.

\begin{remark}
	On the one hand, we choose $\tau$ that minimizes the cost function $\mathcal{J}_p$ discussed later in this subsection. On the other hand, the selection of $\tau$ should help to reduce the proportion of ``false positives" and ''false negatives".
\end{remark}

\begin{figure}[htbp!]
	\centering
	\includegraphics[width=3.5in]{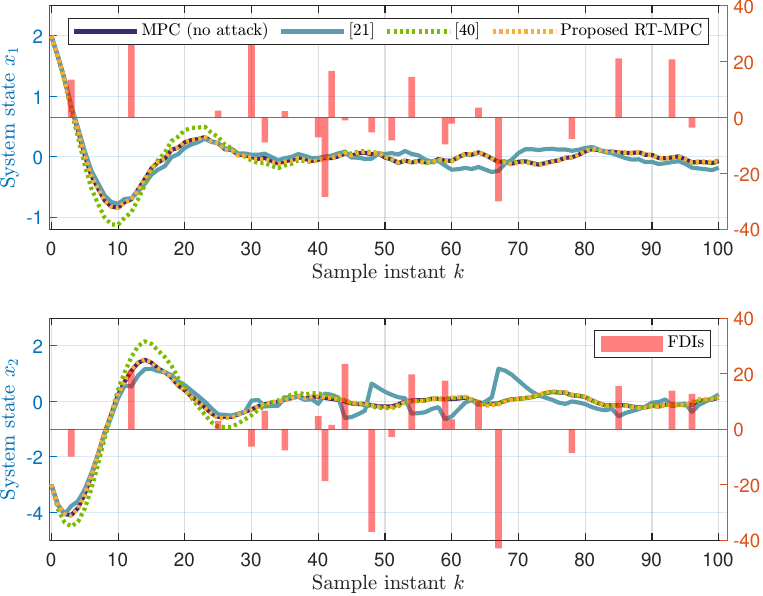}
	\caption{Performance comparison of diverse MPC schemes. The purple curve shows the nominal MPC trajectory \cite{MayneE2000} without attacks, acting as the baseline scenario; the blue curve shows the tube-based MPC scheme in \cite{Mayne2005}, acting as the benchmark method; the dotted green curve shows the resilient MPC scheme in \cite{Tian2023} as comparison; the orange dotted curve shows our proposed resilient scheme. FDI attacks are represented by red bars.}
	\label{Compare}
\end{figure} 

Concerning the system state shown in Fig. \ref{Compare}, we observe the control performance of 1) nonlinear MPC \cite{MayneE2000}  without FDI attacks, 2) tube-based MPC in \cite{Mayne2005}, 3) resilient MPC in \cite{Tian2023} and 4) resilient tube-based MPC (proposed RT-MPC). All the methods are tested under process disturbances $w_1$ and $w_2$ in system states $x_1$ and $x_2$ respectively. The value of the FDI attacks can be seen in Fig. \ref{Compare}, which is greater than the process noises by more than two orders of magnitude.

By comparing the curves, it is evident that our proposed resilient scheme utilizing the attack detection and buffer is superior in resisting FDI attacks on the S-C channel with the presence of bounded disturbances. Its performance infinitely approaches the effect of the baseline non-attack scenario.

To quantitatively demonstrate the superiority of the proposed resilient scheme, we define a cost function as the performance index
$$\mathcal{J}_p = \frac{\sum_{t_k=1}^{N_{sim}}x^TQx + u^TRu}{N_{sim}}.$$

By conducting 100 Monte Carlo experiments, we obtained the results listed in Table \ref{table: comparison}.
The results presented in the table demonstrate that the proposed resilient tube-based MPC scheme outperforms the benchmark scheme (TMPC) with a significantly lower average cost. Compared to the resilient method in \cite{Tian2023}, our approach shows superior performance metrics across a variety of attack scenarios. Specifically, the cost of the resilient scheme is at least 71.80\% lower than that of the non-resilient scheme and at least 15.67\% lower than that of the comparison method. Additionally, the average detection accuracy (\textit{Acc.}) exceeds 99.80\% across diverse scenarios.

\begin{table}[ht]
	\centering
	\setlength{\tabcolsep}{1.5mm}
	\caption{Comparison of Performance Indexes of Diverse Methods under Different Attack Scenarios}
	\label{table: comparison}
	\begin{tabular}{r|ccc}
		\toprule
		\textbf{Metric}  & \textbf{TMPC \cite{Mayne2005}} & \textbf{RMPC \cite{Tian2023}} & \textbf{Proposed RT-MPC } \\ 
		\midrule
		$\bar{\mathcal{J}_p}$ ($\bar{\mathfrak{a}}=0.2, \sigma = 20$)& 7.5354  & 1.6550 & 1.1906 \\ 
		\textit{Acc.} ($\bar{\mathfrak{a}}=0.2, \sigma = 20$) & \textbackslash  & \textbackslash & 99.80\% \\
		\midrule
		$\bar{\mathcal{J}_p}$ ($\bar{\mathfrak{a}}=0.1, \sigma = 20$)& 4.1957  & 1.6528 & 1.1832 \\ 
		\textit{Acc.} ($\bar{\mathfrak{a}}=0.1, \sigma = 20$) & \textbackslash  & \textbackslash & 99.92\% \\
		\midrule
		$\bar{\mathcal{J}_p}$ ($\bar{\mathfrak{a}}=0.1, \sigma = 50$)& 8.6740  & 1.4730 & 1.2105 \\ 
		\textit{Acc.} ($\bar{\mathfrak{a}}=0.1, \sigma = 50$) & \textbackslash  & \textbackslash & 99.97\% \\
		\midrule
		$\bar{\mathcal{J}_p}$ ($\bar{\mathfrak{a}}=0.2, \sigma = 50$)& 16.6919  & 1.4810 & 1.2490 \\ 
		\textit{Acc.} ($\bar{\mathfrak{a}}=0.2, \sigma = 50$) & \textbackslash  & \textbackslash & 99.88\% \\
		\bottomrule
	\end{tabular}
\end{table}

\begin{figure}[ht]
	\centering
	\includegraphics[width=3.2in]{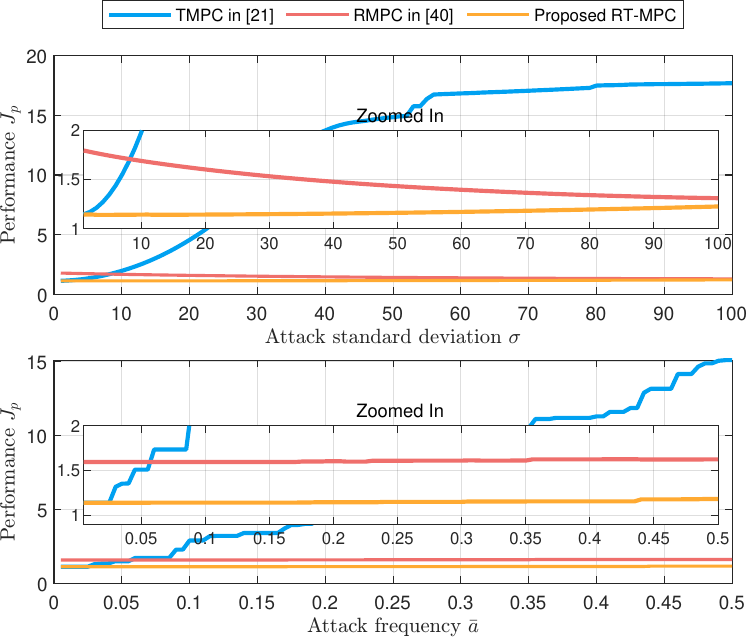}
	\caption{Comparative performance of control schemes under different attack scenarios. Variations in attack scenarios are quantified by the standard deviation $\sigma$ and frequency of attacks $\bar{\mathfrak{a}}$. The blue curve represents the robust control benchmark scheme (TMPC in [21]). The red curve represents the resilient MPC scheme in \cite{Tian2023} used for comparison. The yellow curve indicates the performance of our proposed RT-MPC scheme. Lower $\mathcal{J}_p$ corresponds to better control effects.}
	\label{Changing_sigma_p}
\end{figure}

We evaluate the performance of our scheme and the comparison method under varying attack levels, where the standard deviation $\sigma$ ranges from 1 to 100. We also assess the impact of different attack frequencies ($\bar{\mathfrak{a}}$ varies from 0.005 to 0.5) on the control performance of the three schemes. The results presented in Fig. \ref{Changing_sigma_p} highlight the superiority of our approach.

\begin{remark}
	While the scheme achieves the highest detection rates ($>99.9\%$) for high-deviation, low-frequency FDI attacks ($\sigma > 10 \bar{w}$, $\bar{\mathfrak{a}} > 0.2$), effectiveness decreases for stealthy attacks or systems with sampling periods < 10ms. Implementation requires balancing security needs with computational capabilities, ideally deployed in control systems with redundant processing resources.
\end{remark}

\subsection{Analysis of the impact of strong disturbances}

We analyze the robustness of our proposed resilient control scheme under strong disturbances. We evaluate the performance of the control system when subjected to random disturbances of varying amplitudes. The disturbances were modeled as random noise uniformly distributed within the interval $[-\bar{w}, +\bar{w}]$.
100 rounds are taken and the bound of the disturbance amplitude $\bar{w}$ is in a range of $0.05 \sim 2.00$. The other settings remain the same as in Section \ref{System model}.
The results presenting the mean value of the control scheme’s performance metrics for each disturbance amplitude are summarized in Table \ref{tab:performance_metrics}.

\begin{remark}
In our analysis, we classify disturbances as "strong" when the bound 
$\bar{w}$ approaches or exceeds the order of magnitude of $\|B\|$. Given that 
$\|B\| = 0.0914$ in our case, disturbances with $\bar{w} \ge 0.1$ are considered strong. Moreover, as noted by \cite{Hou2023}, a disturbance bound up to $\bar{w} = 2$ has been used in robustness analysis to test schemes under extreme conditions. Based on these considerations, we select the range $0.05 \sim 2.00$ for our robustness analysis in this subsection. This range allows us to adequately capture both moderate and strong disturbances, including extreme cases.
\end{remark}

\begin{table}[ht]
	\caption{Performance Metrics under Different Disturbance Amplitudes (partial)}
	\label{tab:performance_metrics}
	\begin{center}
		\begin{tabularx}{8cm}{>{\centering\arraybackslash}X >{\centering\arraybackslash}X >{\centering\arraybackslash}X >{\centering\arraybackslash}X >{\centering\arraybackslash}X}
			\toprule
			\textbf{Disturbance bound $\bar{w}$} & \textbf{\textit{Acc.(\%)}} & \textbf{$\mathcal{J}_p$} & \textbf{\textit{Saving(\%)}} & \textbf{\textit{Tracking error(\%)}} \\
			\midrule
			0.05 & 99.95 & 1.1790 & 78.82 & 1.00 \\
			\midrule
			0.1 & 99.84 & 1.2587 & 78.61 & 1.47 \\
			\midrule
			0.2 & 99.72 & 1.5308 & 75.43 & 1.81 \\
			\midrule
			0.5 & 99.25 & 3.3942 & 59.69 & 1.74 \\
			\midrule
			1.0 & 98.45 & 10.0748 & 33.89 & 1.49 \\
			\midrule
			2.0 & 93.91 & 37.2121 & 12.36 & 2.31 \\
			\bottomrule\\
		\end{tabularx}
	\end{center}
	Note: All data presented are averages from 100 Monte Carlo simulations. \textbf{\textit{Acc.}} -- attack detection accuracy; \textbf{$\mathcal{J}_p$} -- cost function;\\
	\textbf{\textit{Saving}} -- cost function compared to TMPC \cite{Mayne2005}; \textbf{\textit{Tracking error}} -- cost function compared to nominal MPC \cite{MayneE2000} without attack.
\end{table}

\begin{figure}[htbp!]
	\centering
	\includegraphics[width=3in]{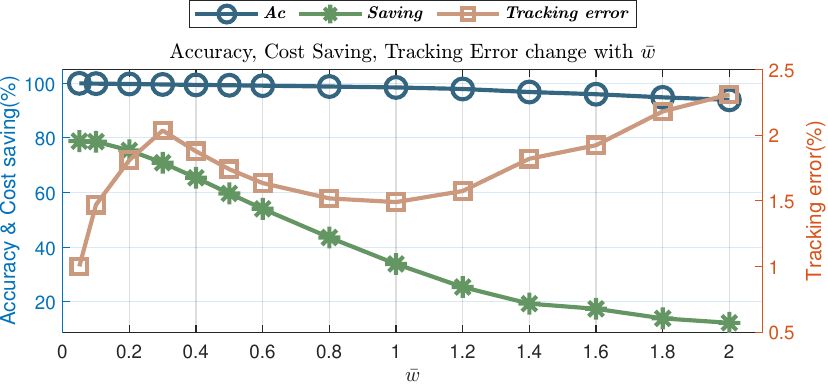}
	\caption{Impact of changing $\bar{w}$ from 0.05 to 2. The detection accuracy and the cost saving keep declining, while the tracking error shows its local minimum around $\bar{w} = 1$.}
	\label{linechart}
\end{figure}

The visible line chart is presented in Fig. \ref{linechart}. The results demonstrate that as the amplitude of disturbances increases, the detection accuracy slightly decreases but remains above 90\%, highlighting the robustness of our detection scheme against significant disturbances. However, cost savings decline from approximately 80\% to 12\% as the disturbance bound increases to 2. This finding suggests that in extreme disturbance scenarios, process disturbances become the primary determinant of performance, rather than attacks. Regarding tracking error, it peaks at $\bar{w} = 0.3$, reaches its minimum at $\bar{w} = 1$, and then steadily increases until $\bar{w} = 2$. These results indicate that resilience against FDI attacks is most effective when $\bar{w}$ is around 1. This validates the scheme's effectiveness in ensuring resilience and robustness under adverse conditions.

\subsection{Analysis of the impact of attack threshold}

\begin{figure}[htbp!]
	\centering
	\includegraphics[width=3in]{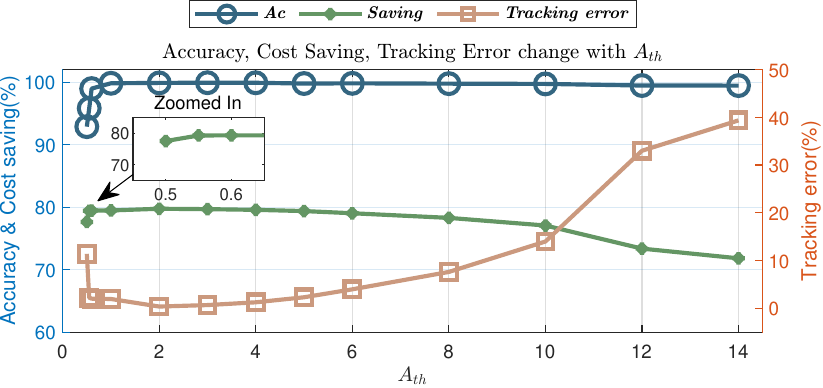}
	\caption{Impact of changing $A_{th}$ from 0.5 to 14. All performance metrics show peak/valley values. The optimal value for $A_{th}$ ranges from 2 to 6.}
	\label{Impact of changing Ath}
\end{figure}

We explore the influence of the attack threshold of the detector on performance metrics. The parameter $A_{th}$ is varied from 0.5 (the order of magnitude of process disturbance) to 14 (the order of magnitude of FDI attack). The outcomes are presented in Table \ref{tab2:performance_metrics} and Fig. \ref{Impact of changing Ath}.

\begin{table}[ht]
	\caption{Performance Metrics under Different Detection Threshold (partial)}
	\label{tab2:performance_metrics}
	\begin{center}
		\begin{tabularx}{8cm}{>{\centering\arraybackslash}X >{\centering\arraybackslash}X >{\centering\arraybackslash}X >{\centering\arraybackslash}X >{\centering\arraybackslash}X}
			\toprule
			\textbf{Detection threshold $A_{th}$} & \textbf{\textit{Acc.(\%)}} & \textbf{$\mathcal{J}_p$} & \textbf{\textit{Saving(\%)}} & \textbf{\textit{Tracking error(\%)}} \\
			\midrule
			0.5 & 92.86 &  1.3058 & 77.62 & 11.38 \\
			\midrule
			1 & 99.80 & 1.1888 & 79.45 & 1.92 \\
			\midrule
			4 & 99.88 & 1.1814 & 79.55 & 1.23 \\
			\midrule
			8 & 99.74 & 1.2555 & 78.26 & 7.59 \\
			\midrule
			14 & 99.43 & 1.6276 & 71.79 & 39.34 \\
			\bottomrule\\
		\end{tabularx}
	\end{center}

\end{table}

Fig.\ref{Impact of changing Ath} indicates that as $A_{th}$ increases from 0.5, both detection accuracy and cost saving initially rise and then decline. When $A_{th}$ reaches approximately 1, the accuracy consistently maintains a high value (exceeding 99\%), while cost saving stabilizes around 80\% within the range of $A_{th}$ from 0.5 to 6, indicating favorable results. However, a notable downward trend in cost saving occurs once $A_{th}$ surpasses 10. The minimum tracking error appears between $A_{th}$ values of 2 and 4 and gradually increases for $A_{th}$ values greater than 6. Thus, the optimal value for $A_{th}$ is in the range of 2 to 6.

\subsection{Resilient control on the HVAC system}\label{Nonlinear Case}
\begin{figure}[htbp!]
	\centering
	\includegraphics[width=3in]{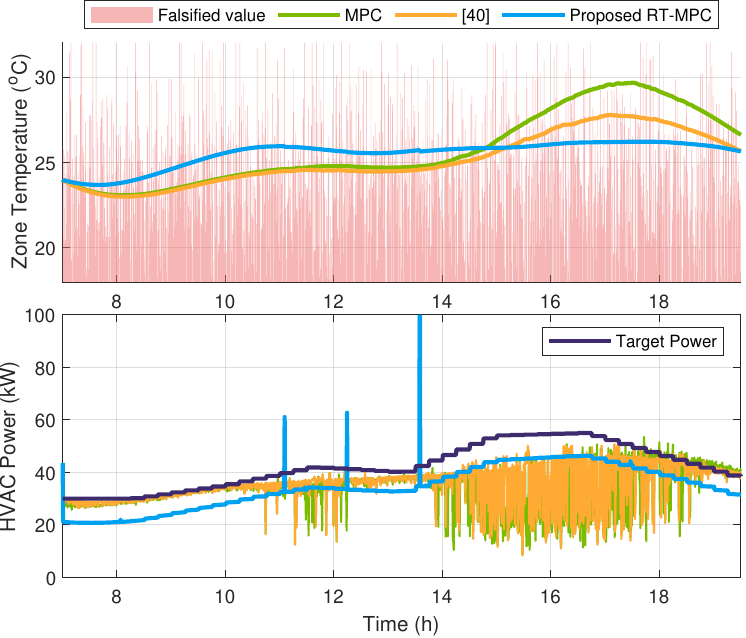}
	\caption{Time-dependent curves of zone temperature and air supply power under different control schemes. The control objective is to track the rated power of the power grid, which varies over 24 hours while maintaining a stable zone temperature. Red bars show the falsified zone temperature; The rated power is indicated by the purple line. The MPC acting as the baseline method is represented by the green line, the comparison scheme \cite{Tian2023} by the orange line, and the proposed RT-MPC by the blue line.}
	\label{HVAC_Tr_P}
\end{figure}

The proposed resilient control scheme is tested using the digital twin models of a smart building with a single-chiller Heating, Ventilation and Air Conditioning (HVAC) system. The results are compared with an existing resilient min-max MPC scheme \cite{Tian2023}. The only difference is that we take the comparison in the amplitude-unbounded scenario. The time span of the simulation is 24 hours. The standard MPC serves as the benchmark method, which works perfectly without FDI attacks but yields large overall power profile deviations when attacks occur. The Root Mean Square Error (RMSE) of the power tracking and the computational time are compared between RMPC in \cite{Tian2023} and our proposed scheme. The results are shown in Table \ref{table:hvac_comparison} and Fig. \ref{HVAC_Tr_P}.

\begin{table}[htbp]
	\centering
	\caption{Comparison of Performance and Computation Time}
	\label{table:hvac_comparison}
	\begin{tabular}{l ccc}
		\toprule
		\textbf{Metric} & \textbf{MPC} & \textbf{RMPC \cite{Tian2023}} & \textbf{Proposed RT-MPC} \\ 
		\midrule
		RMSE (Power Tracking) & 15.2897 & 11.7480 & 8.7633  \\ 
		Mean Time (sec/step) & 0.8347 & 2.1499 & 1.0284 \\
		\bottomrule
	\end{tabular}
\end{table}

It shows that our method not only reduces the power tracking error by over 25.4\% against the method in \cite{Tian2023} but also shows higher robustness against higher levels of attack magnitudes. Moreover, the computational requirement of solving the proposed resilient control is sufficiently efficient for real-time applications, as shown in Table \ref{table:hvac_comparison}.

\subsection{Nonlinear case}
To validate the practical applicability of our proposed resilient control scheme for mostly nonlinear CPSs, we consider the following control case of a 3-dimensional nonlinear system. We control the position and attitude of a non-holonomic vehicle within a plane. Its dynamic equation is modeled as follows:
\begin{equation}
	\frac{\mathrm{d}}{\mathrm{d}t}\begin{bmatrix}p_x\\p_y\\\theta\end{bmatrix}=\begin{bmatrix}\cos\theta&&0\\\sin\theta&&0\\0&&1\end{bmatrix}\begin{bmatrix}v\\\omega\end{bmatrix},
	\end{equation}
where $p_x$ and $p_y$ are the coordinates of the vehicle on X and Y-axis, $\theta$ is the angle between the positive X-axis direction. These three variables constitute the 3-dimensional state vector $x = [p_x, p_y, \theta]^\top$. The control input is $u = [v, \omega]^\top$, representing the linear velocity and angular velocity. This experiment aligns with recent studies that have successfully stabilized a 3-dimensional vehicle using advanced control methods\cite{NlCarModel}.

In this test case, we set the prediction horizon to $N = 20$, the constraint condition to
(\ref{eq:X_Constraint}), (\ref{eq:U_Constraint}) and (\ref{eq:W_Constraint}) 
where $\overline{x} = - \underline{x} =[10, 10, \pi]^\top$, $\overline{u} = - \underline{u} =[0.5, 0.1]^\top$, $\overline{w} = - \underline{w} =[0.1, 0.1, 0.03]^\top$.
The cost matrix is set to $Q=0.1I_3$ and $R=0.05I_2$. The FDI attacks are recognized as $N_\mathcal{a} \sim B(100,0.1)$ and $\mathcal{a}_k \sim N(0, 0.45^2)$. The hyperparameter $d_{th} = 1.5$ with $\tau = 5.8$.  All other settings remain unchanged compared to above linear two-dimensional case. The initial conditions for the vehicle are set to $[-5\mathrm{~}4\mathrm{~}-\frac{\pi}{2}]^\top$ with the control objective of $[0\mathrm{~} 0 \mathrm{~} 0]^\top$. 
When addressing nonlinear control problems, our approach updates the linearized system equation $A$ and input equation $B$ prior to each prediction step. For this continuous-time system, we discretize it at a sampling period of $T_s = 0.1s$ using a zero-order holder to facilitate computer control.
\begin{figure}[h]
	\centering
	\includegraphics[width=3.5in]{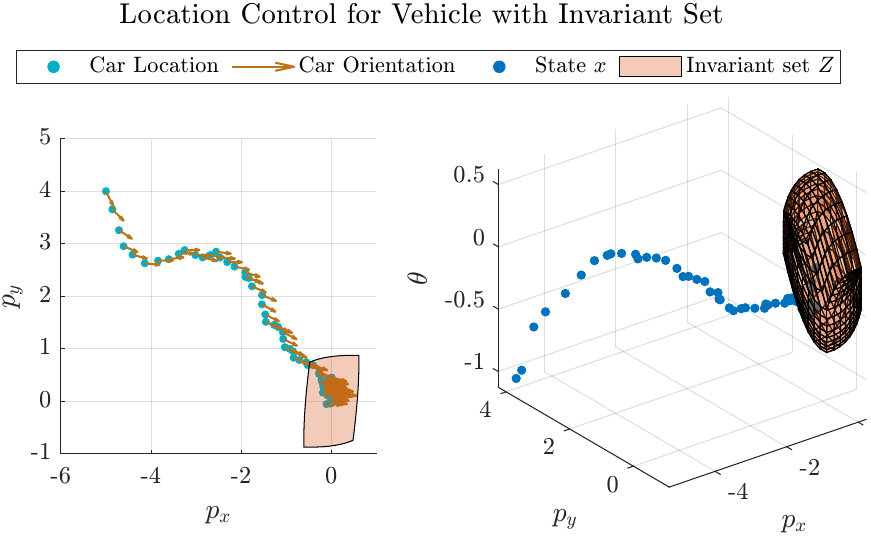}
	\caption{State trajectories and invariant set for vehicle position control in both dot-arrow style and 3-D style. The solid green points represent the coordinates of the vehicle in the plane, and the arrows indicate the orientation of the car. The orange region and the orange polytope show the invariant set $Z$ for disturbances and FDI attacks. The blue dots represent the trajectory of state variables in space.}
	\label{LO_Z}
\end{figure}
The vehicle's trajectory is depicted in Fig. \ref{LO_Z}, where the solid blue points represent the coordinates of the vehicle in the plane, and the arrows indicate the orientation of the vehicle. The light orange region shows the invariant set $Z$ for disturbances and FDI attacks.

The results demonstrate that our scheme can be generalized to network security control against FDI attacks in 3-dimensional and simplified nonlinear systems. This implies that our resilient scheme is practically significant for high-dimensional, nonlinear complex systems in the real world.

As for strong nonlinear cases, drawing upon the work of \cite{Zhao2024}, it is feasible to approximate nonlinear CPSs with linear models, which can then be integrated with our approach to enhance its effectiveness.
	\section{Conclusions}

This paper presents a resilient MPC algorithm to address the issue of CPSs experiencing amplitude-unbounded FDI attacks in the S-C channel. The proposed countermeasure has two key features.
﻿First, it utilizes the set-theoretic tube method. This method guarantees the input-to-state stability of systems under bounded disturbances and FDI attacks. The feasibility of the proposed buffering technique is proven through probability theory.
Second, it employs a resilient mechanism based on attack detection and sequence buffering. This mechanism leverages the inherent characteristics of the rolling optimization method to effectively identify and mitigate the impact of the unbounded attacks.
A crucial aspect of this countermeasure is that attack identification and control law selection are performed entirely within the resilient tube-based MPC controller. This process is independent of the affected sensor-controller channel.
Experimental testing of the proposed algorithm on different scenarios of CPSs demonstrates the superior performance of the resilient scheme compared to the existing methods.
In future research, the focus will be on enhancing the universality of the resilient MPC algorithm. This will involve considering multi-channel attack resistance and data-driven attack model identification.

	\vspace{11pt}
	
	
	\vspace{11pt}
	

	\bibliographystyle{unsrt}

	\begin{IEEEbiography}[{\includegraphics[width=1in,height=1.25in,clip,keepaspectratio]{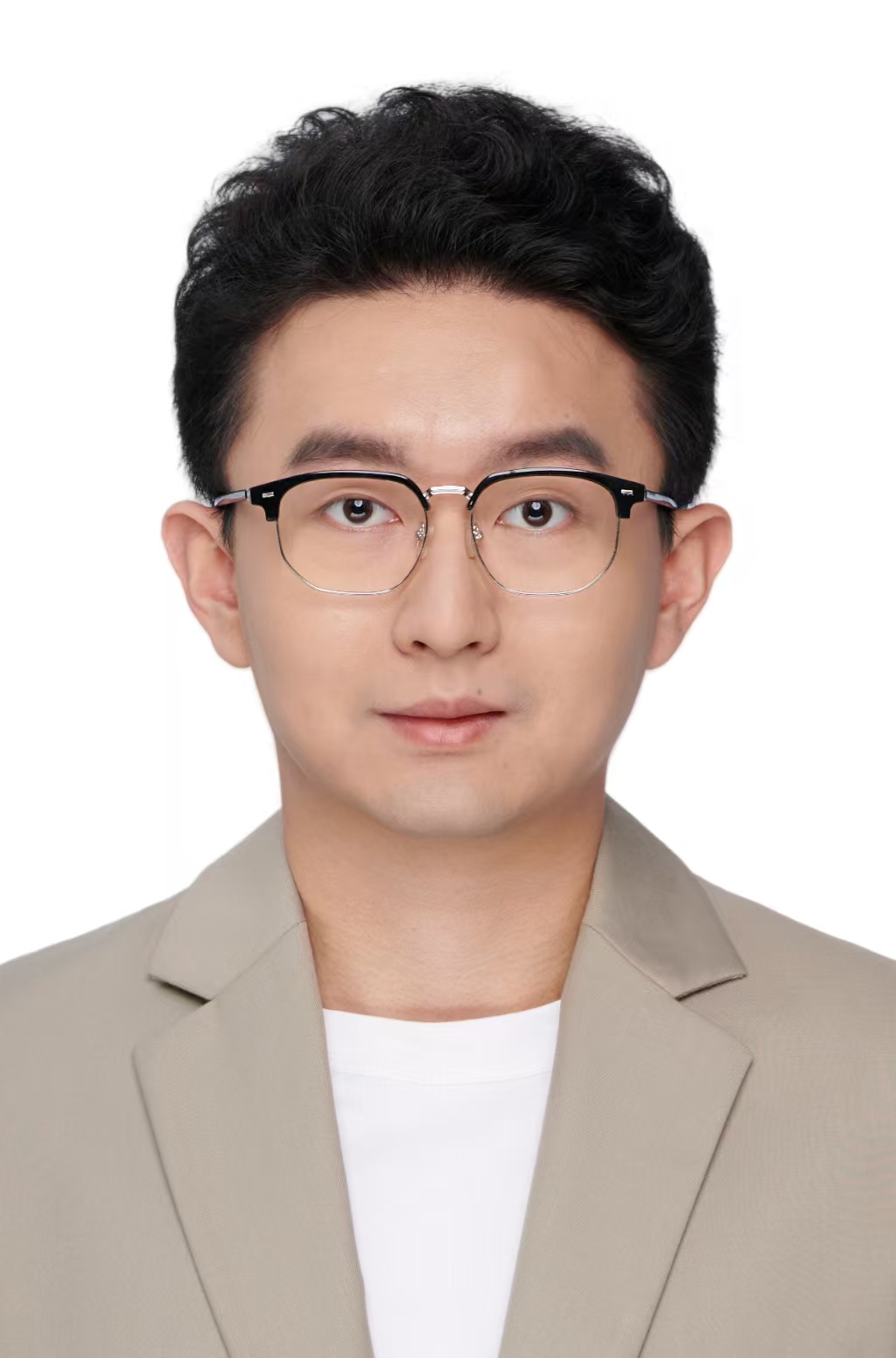}}]{Yuzhou Xiao}
		  received the B.S. degree in Automation from the Beijing Institute of Technology, Beijing, China, in 2023. He is currently pursuing the M.S. degree in control science and engineering with the Beijing Institute of Technology, Beijing, China. His current research interests include robust model predictive control and resilient schemes in cybersecurity.
	\end{IEEEbiography}
	
	\begin{IEEEbiography}[{\includegraphics[width=1in,height=1.25in,clip,keepaspectratio]{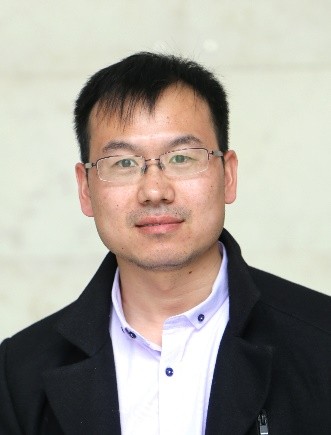}}]{Senchun Chai}
		 received the B.S. and master’s degrees in control science and engineering from the Beijing Institute of Technology, Beijing, China, in 2001 and 2004, respectively, and the Ph.D. degree in networked control system from the University of South Wales, Pontypridd, U.K., in 2007. He is currently a Professor with the School of Automation, Beijing Institute of Technology. He was a Research Fellow with Cranfield University, U.K., from 2009 to 2010, and a Visiting Scholar with the University of Illinois at Urbana–Champaign, Urbana, USA, from January 2010 to May 2010. He has published over 100 journals and conference papers. His current research interests include flight control systems, networked control systems, embedded systems, and multi-agent control systems.
	\end{IEEEbiography}
	
	\begin{IEEEbiography}[{\includegraphics[width=1in,height=1.25in,clip,keepaspectratio]{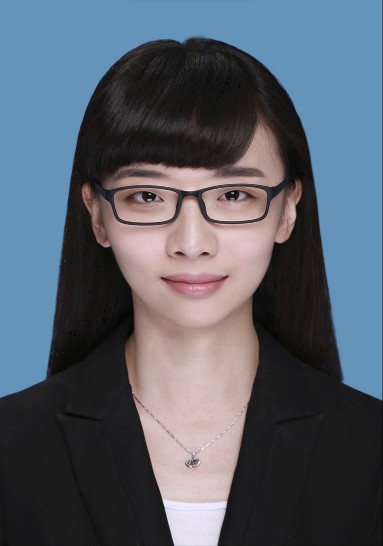}}]{Li Dai}
	  received the B.S. degree in information and computing science and the Ph.D. degree in control science and engineering from the Beijing Institute of Technology, Beijing, China, in 2010 and 2016, respectively. She is currently a Professor with the School of Automation, Beijing Institute of Technology. Her research interests include model predictive control, distributed control, data-driven control, stochastic systems, and networked control systems.
	\end{IEEEbiography}
	
	\begin{IEEEbiography}[{\includegraphics[width=1in,height=1.25in,clip,keepaspectratio]{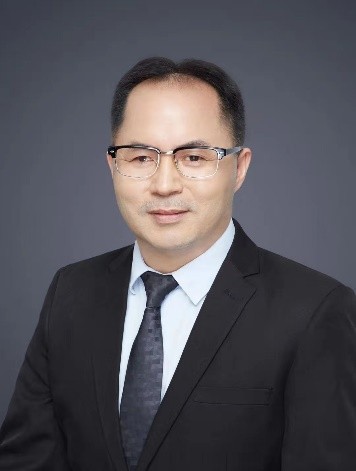}}]{Yuanqing Xia}
		 received the Ph.D. degree in control theory and control engineering from Beihang University, Beijing, China, in 2001. He was a research fellow in several academic institutions during 2002 to 2008, including China Academy of Sciences, National University of Singapore, University of Glamorgan, Innsbruck Medical University (Austria), etc. Since 2004, he has been with School of Automation, Beijing Institute of Technology, Beijing, China, where he is currently a Professor. He obtained the National Outstanding Youth Foundation of China in 2012, and was honored as a Yangtze River Scholar Distinguished Professor in 2016. His research interests include cloud control systems, networked control systems, signal processing, active disturbance rejection control, unmanned system control and flight control.
	\end{IEEEbiography}
	
	\begin{IEEEbiography}[{\includegraphics[width=1in,height=1.25in,clip,keepaspectratio]{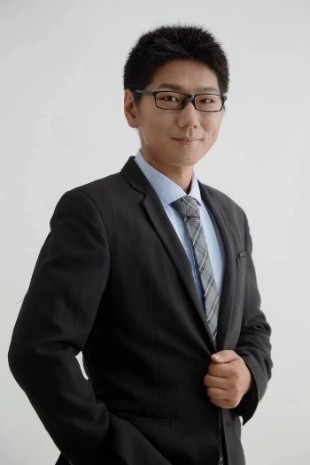}}]{Runqi Chai}
		 received the B.S. degree in information and computing science from the North China University of Technology, Beijing, China, in 2015, and the Ph.D. degree in aerospace engineering from Cranfield University, Cranfield, U.K., in August 2018. He is currently a Professor with the School of Automation, Beijing Institute of Technology, Beijing. He was a Research Fellow with Cranfield University, from 2018 to 2022. His research interests include trajectory optimization, networked control systems, and multiagent control systems.
	\end{IEEEbiography}

	\vfill
	
\end{document}